# Application of polynomial texture mapping in process of digitalization of cultural heritage

Authors:

B. Malešević[1], R. Obradović[2], B. Banjac[1,2], I. Jovović[1], M. Makragić[1]

[1] Faculty of Electrical Engineering, University of Belgrade
[2] Faculty of Technical Sciences, University of Novi Sad

Abstract

In this paper we present modern texture mapping techniques and several applications of polynomial texture mapping in cultural heritage programs. We also consider some well-known and some new methods for mathematical procedure that is involved in generation of polynomial texture maps.

1. Introduction to texture mapping

Texture mapping is technique in computer graphic of applying images to surfaces. This technique allows greater realism of computer generated images. During last thirty years many different algorithms for texture mapping were developed. These algorithms are also connected to other algorithms for signal filtering to reduce effects like aliasing. Some of them never had much of practical application because of hardware limitations. Most of API-s for 3D graphics implemented several methods for texture mapping, but as programming graphical card was pretty complicated and demanded detailed knowledge of specific architecture for every graphical card, not many non-standard techniques had their implementations other than those in rendering software which relay on central processing unit instead of graphical card. With coming of programmable pixel shader processors, those limitations were lifted. That allowed for software developers to implement mapping and filtering algorithms that they thought appropriate for their needs.

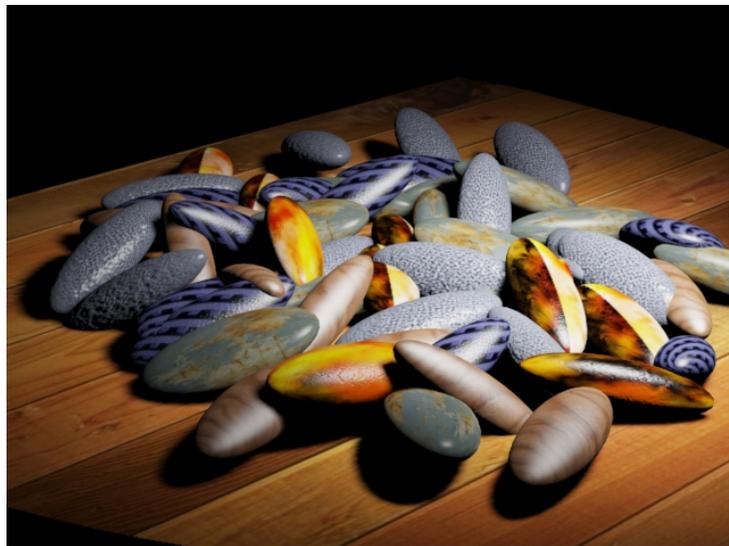

Figure 1. Computer generated image with modern texture mapping techniques applied for greater realism

One of standard methods for texture mapping involves simple "gluing" of image to polygon. This simply changes color of pixels at polygon to given value. As today graphical hardware allows very fast application of this method, this is widely used method. With additional filtering this method allows for decent representation of smooth surfaces (like processed marble), but rough surfaces will look flat. Because shading is done in same way to whole textured area, this creates unrealistic lightning effects, especially if lightning source is moving, and shadows on surface don't change [6]. This method with use of appropriate filters and lightning is still one of most used.

Another approach is by use of multilayer textures. One of textures is used to define color of each pixel. Other texture has purpose of defining surface to which colors are applied. Instead of using more complex surface which has to be approximated by larger number of polygons, by calculating lightning as if surface is more complex, better shadowing effects can be achieved. Second texture is used in several ways. First is that in surface texture is contained height information for each pixel. This is usually called bump mapping. Surface texture can also contain normal of each pixel, coded as RGB values. While these kinds of mapping will not allow self-shadowing, and will not affect shape of shadows, they do allow greater realism at very low hardware cost. To allow even greater realism, displacement mapping is used. Displacement mapping creates new polygons from existing polygon using displacement map. This creates very realistic surfaces, because of use of greater number of polygons. Unfortunately this also comes with much larger hardware cost [7].

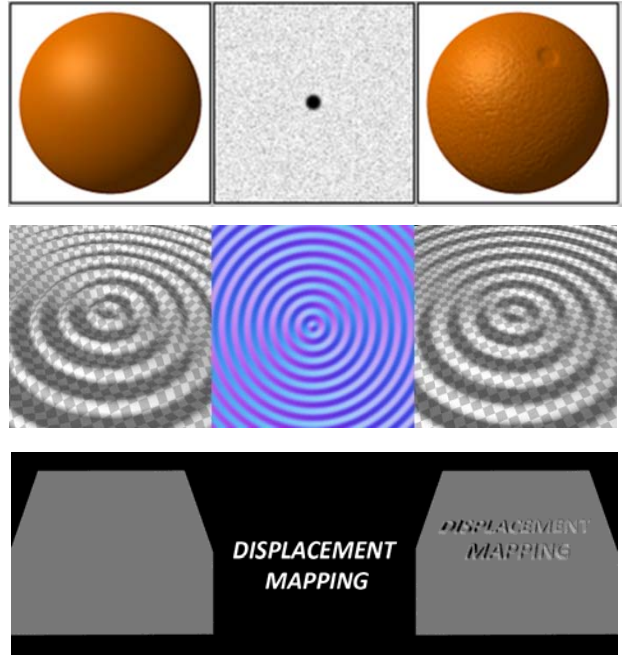

Figure 2.
Upper image: Sphere, bump map, and sphere with bump map applied
Middle image: Surface with 19 600 polygons, normal map generated from it, and plane with normal mapping applied
Lower image: Plane of 22500 polygons before and after displacement mapping was applied

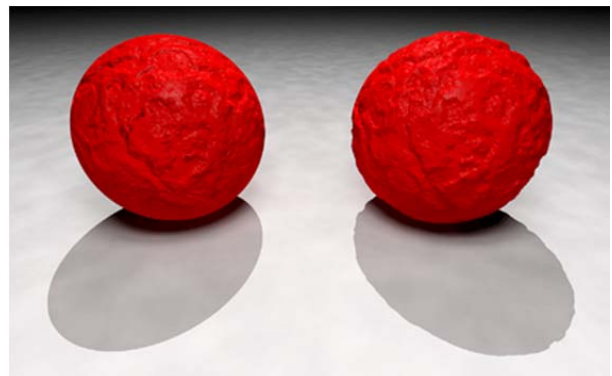

Figure 3. Left sphere has bump map applied, right sphere has displacement map applied

With all of these methods it is very difficult to create surface texture. In most cases these surfaces are modeled by some computer modeling software, and later converted to texture data. There were also several attempts to create some method of automatic texture building algorithms from image. Often these attempts didn't have desired effect because of complexity of task. While equipment for 3D scanning of objects does exist, it is fairly expensive and rare. Because of that some other method for creating images to realistic textures were developed.

2. Polynomial texture mapping

Polynomial texture mapping is method for texture mapping that uses very realistic model for shadows changing dependent of light source position [1-10]. This method is based on some older techniques that were working on dependency of color of texture and position of light source. One of first techniques for this is The Bidirectional Reflectance Distribution Function [9]. For this texture method position of observer $(\alpha_o, \beta_o)$ and position of light source $(\alpha_l, \beta_l)$ as well as wavelength of light $\lambda$ are taken account in function $BRDF(\alpha_o, \beta_o, \alpha_l, \beta_l, \lambda)$ which calculates intensity of light reflectance to observer. This method gives much better results than any of previously described because it allows for self-shadowing and much more realistic lightning behavior. Bidirectional texture function uses previous function and allows it to vary across surface. To reduce number of factors, wavelengths are limited to standard Red, Green and Blue values. Function of $(\alpha_o, \beta_o, \alpha_l, \beta_l, u, v)$, where $u$ and $v$ are position of particular pixel on texture space, as previous gives reflectance from light source. While this method allowed for very good results, it was very complicated to work with. To get correct results position of light source and observer camera had to be very carefully calibrated. Because of nature of procedure, very large number of values had to be sampled to give correct results, so it is difficult to build these kind of textures.

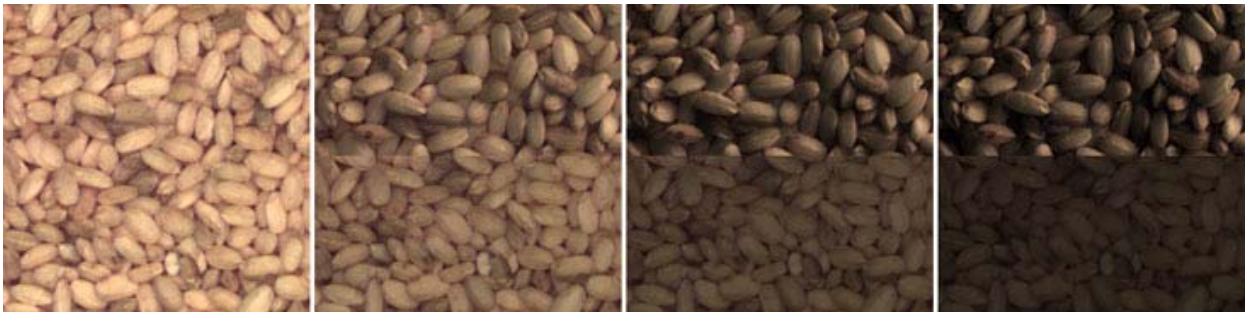

Figure 4. Upper half of image has Polynomial texture map applied, lower half of image has only simple bitmap (source [1])

Polynomial texture maps were developed to avoid pitfalls of previous methods. First limitation that was introduced is that observer's position is fixated. As observers position during photographing is fixed, only changing factor is position of lightning. With that some effects that are tied to observer's position, like specular properties of material, are overseen but diffuse lightning and shadowing details are still kept correct. When using this kind of textures, it is presumed that during the acquisition of data all specular properties were removed, either by use of correct materials or polarized light.

For generation of polynomial texture maps several solutions are invented. On field often used is simple cupola made of wire. By fixing camera on top of cupola and positioning proper light source in adequate number of positions on cupola, after capturing large number of images, polynomial texture map can be generated. In laboratory conditions several institutions have made cupola of non-reflective materials. On those cupolas light sources are mounted on regular intervals, and opening for camera is positioned on top. Some research facilities even developed apparatus for gathering of forensic data, and are being tested for field use.

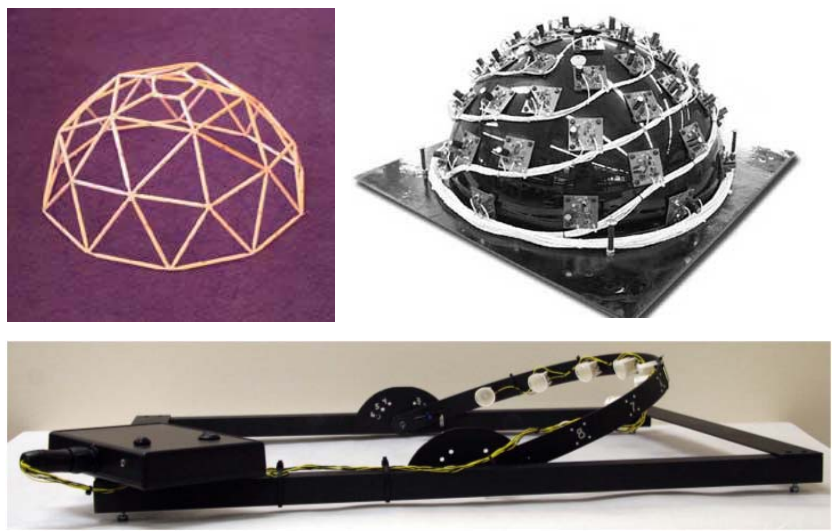

Figure 5. Several different apparatus for generating of Polynomial texture maps

Initial collection of images is followed by calculations of polynomial coefficients that are done for each color of each pixel in images. After calculation we can determine color of each pixel by application of formulas

$$C_r(u,v,l_u,l_v) = a_{r_0}(u,v) \cdot l_u^2 + a_{r_1}(u,v) \cdot l_v^2 + a_{r_2}(u,v) \cdot l_u \cdot l_v + a_{r_3}(u,v) \cdot l_u + a_{r_4}(u,v) \cdot l_v + a_{r_5}(u,v)$$

$$C_g(u,v,l_u,l_v) = a_{g_0}(u,v) \cdot l_u^2 + a_{g_1}(u,v) \cdot l_v^2 + a_{g_2}(u,v) \cdot l_u \cdot l_v + a_{g_3}(u,v) \cdot l_u + a_{g_4}(u,v) \cdot l_v + a_{g_5}(u,v)$$

$$C_b(u,v,l_u,l_v) = a_{b_0}(u,v) \cdot l_u^2 + a_{b_1}(u,v) \cdot l_v^2 + a_{b_2}(u,v) \cdot l_u \cdot l_v + a_{b_3}(u,v) \cdot l_u + a_{b_4}(u,v) \cdot l_v + a_{b_5}(u,v)$$

where $(u,v)$ define position of pixel in local texture coordinate space, and $(l_u, l_v)$ define projections of normalized light vector. After initial polynomial fitting coefficients $a_{r_0} - a_{r_5}, a_{g_0} - a_{g_5}, a_{b_0} - a_{b_5}$ are stored, and only upper formulas are used. This allows for great speed of calculation, and great practical use of this type of texturing. Here should be mentioned that after fitting some values are mellowed, so for example some shadow that were extremely hard shall be little mellower. However this does not have effect of spatial blurring, but only on light reflecting. As it is often that light reflectance for each color behaves similarly, previous calculation can be generalized as:

$$L(u,v,l_u,l_v) = a_0(u,v) \cdot l_u^2 + a_1(u,v) \cdot l_v^2 + a_2(u,v) \cdot l_u \cdot l_v + a_3(u,v) \cdot l_u + a_4(u,v) \cdot l_v + a_5(u,v)$$

$$C_r(u,v) = R(u,v) \cdot L(u,v)$$

$$C_g(u,v) = G(u,v) \cdot L(u,v)$$

$$C_b(u,v) = B(u,v) \cdot L(u,v)$$

where $L$ is luminescence at point and $R, G, B$ color factors. This type of storing takes half less space for storage, and number of calculations done while displaying texture are reduced to one third, but asks for larger initial calculation. One of additional ways to reduce used space for storage of texture is to use scale $\lambda$ and bias $\Omega$. With use of this factor instead of storing 6 integers with 32 or 64 bites of data for coefficients, we reduce them to 8 bit integer, and later reconstruct value as $a_i = \lambda_i \cdot (a_i' + \Omega_i)$, where scale and bias are assigned on level of whole texture.

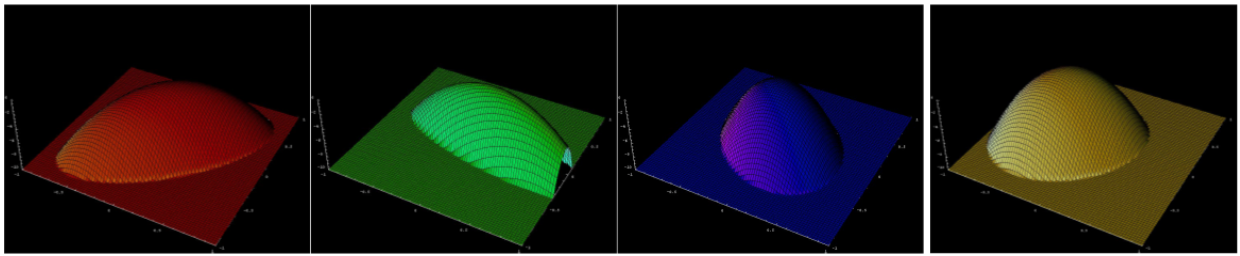

Figure 6. Graphical representation of fitted polynomial surface for color red, green, blue and luminescence value

Polynomial fitting is done by solving system of equations by $a_0, a_1, \ldots, a_5$ given in the matrix form as [1]:

$$\begin{bmatrix} l_{u_0}^2 & l_{v_0}^2 & l_{u_0} \cdot l_{v_0} & l_{u_0} & l_{v_0} & 1 \\ l_{u_1}^2 & l_{v_1}^2 & l_{u_1} \cdot l_{v_1} & l_{u_1} & l_{v_1} & 1 \\ \vdots & \vdots & \vdots & \vdots & \vdots & \vdots \\ l_{u_{N-1}}^2 & l_{v_{N-1}}^2 & l_{u_{N-1}} \cdot l_{v_{N-1}} & l_{u_{N-1}} & l_{v_{N-1}} & 1 \end{bmatrix} \cdot \begin{bmatrix} a_0 \\ a_1 \\ a_2 \\ a_3 \\ a_4 \\ a_5 \end{bmatrix} = \begin{bmatrix} L_0 \\ L_1 \\ L_2 \\ \vdots \\ L_{N-1} \end{bmatrix}$$

Standard approach to considering the system is the least-squares solution using Singular Value Decomposition [1], [15]. Previous system can be written in the following vector form $\Lambda \cdot \vec{a} = \vec{L}$. If the system $\Lambda \cdot \vec{a} = \vec{L}$ has a solution, general method for solving it is presented in papers [11], [12]. If the number of rows is $N = 6$ and if the rank of matrix $\Lambda$ is $\rho = 6$, then the matrix $\Lambda$ is invertible and $\vec{a} = \Lambda^{-1} \cdot \vec{L}$. In other cases $\rho \neq 6$ or $N \neq 6$ the general solution can be determined in form $\vec{a} = \Lambda^{(1)} \cdot \vec{L}$ where $\Lambda^{(1)}$ is the general {1}-inverse determined in the Rohde's general form [11], [12]:

$$\Lambda^{(1)} = P \cdot \begin{bmatrix} I_\rho & U \\ V & W \end{bmatrix} \cdot Q$$

where $U = [u_{ij}]$, $V = [v_{ij}]$ and $W = [w_{ij}]$ are arbitrary matrices of corresponding dimensions and $P$ and $Q$ are regular such that:

$$Q \cdot A \cdot P = E_\rho = \begin{bmatrix} I_\rho & 0 \\ 0 & 0 \end{bmatrix},$$

where $I_\rho$ is $a \times a$ identity matrix. If the system $\Lambda \cdot \vec{a} = \vec{L}$ has no solution, then we can look for pseudo-solutions in the form $\vec{a} = \Lambda^{(1)} \cdot \vec{L}$, see [13], [15]. In practice, the {1}-inverse $\Lambda^{(1)}$ is chosen as the Moore--Penrose pseudoinverse $\Lambda^+$ by which we obtain standard the least-squares solution $\vec{a} = \Lambda^+ \cdot \vec{L}$ (see Section "What kind of answer is $\Lambda^+ \cdot \vec{b}$ ?" in the Chapter 2 of the book [13]).

3. Application of polynomial texture mapping

Polynomial texture mapping is considerably young technology, but it has already found many applications. Like most of its predecessors it has found some applications in gaming industry and image rendering software. By use of polynomial texture mapping objects that were very complex, with use of many vertices can be represented by much simpler shape. One of development's that allowed this technology are programmable shader processors that are now common in graphic cards.

Another application that was found for this type of texture mapping is in two-dimensional graphics. While ordinary images can contain much information about object that is photographed, polynomial texture mapping allows further contrast enchantments. This technology has already been applied on some institutes to preserve digital copy of significant works of art, or historical artifacts [8]. As user have good approximation of surface behavior under different light, many details that were previously inaccessible are now preserved. On old paintings by use of low light can be observed painters brush stroke. On direct light observers can see work of art, in conditions that are not allowed in most galleries.

With old tablets or imprints change of light can enhance contrast of some details [5],[13]. This was already recognized by some governments and studies of application of polynomial texture mapping in crime forensic science are in progress. Currently it has shown great progress and further applications may be available in future.

While lightning manipulation of flat surfaces allows much better understanding of cultural heritage, another application is available. By calculating direction in which light reflectance is at maximum, normal of each pixel can be calculated as is shown in following inequality [4]

$$l_{u0} = \frac{a_2 \cdot a_4 - 2 \cdot a_1 . a_3}{4 \cdot a_0 \cdot a_1 - a_2{}^2}$$

$$l_{v0} = \frac{a_2 \cdot a_3 - 2 \cdot a_0 . a_4}{4 \cdot a_0 \cdot a_1 - a_2{}^2}$$

$$\bar{N} = \left(l_{u0}, l_{v0}, \sqrt{1 - l_{u0}{}^2 - l_{v0}{}^2}\right)$$

This allows for forming of three dimensional model of observed object. While this method is very similar to computer vision method photometric stereo, it does not need specular properties of material to be well determined, and allows much more freedom with objects that consist of several materials. Here should also be mentioned that three dimensional models created by this method are even more precise that those done by laser measuring.

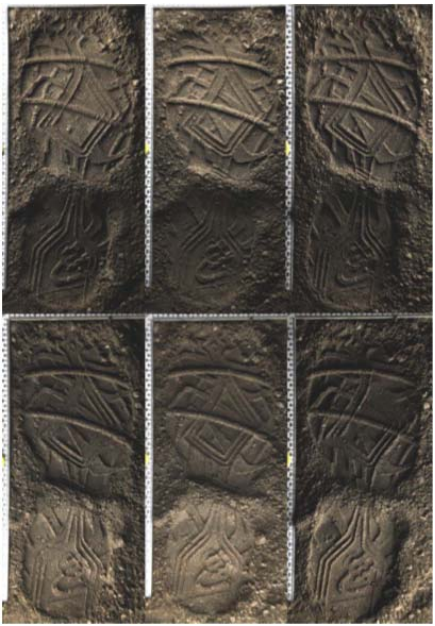

Figure 7. Image of forensic evidence collected as polynomial texture map and later processed (source [10])

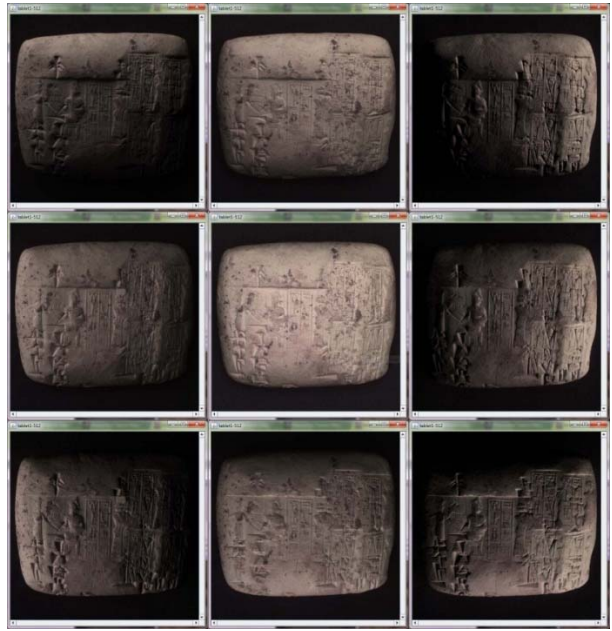

Figure 8. Stone tablet digitalized as polynomial texture map (source [1])

4. Conclusion

In this paper is presented basis for one process of digitalization of cultural heritage such as small archaeological digs, frescos, paintings, etc. Presented polynomial texture mapping is mathematically analyzed with one new method for solving linear systems. In further research it will be analyzed several different forms of generalized {1}-inverse of matrix $\Lambda$ in order to study the various accurate and approximate solutions [14], [15]. At the end, let us emphasize that described method for generation of polynomial texture maps can be realized with very simple hardware, and with very little investment, which could significantly contribute digitalization of cultural treasure.

5. Acknowledgement

Research is partially supported by the Ministry of Science and Education of the Republic of Serbia, Grant No. III 44006 and ON 174032.

Literature:

[1] T. Malzbender, D. Gelb, H. Wolters: *Polynomial texture maps*, Proceedings of SIGGRAPH 2001 conference, pp. 519-528, 2001. (http://www.hpl.hp.com/research/ptm/)

[2] Ø. Hammer, S. Bengtson, T. Malzbender, D. Gelb: *Imaging fossils using reflectance transformation and interactive manipulation of virtual light sources*, Palaeontologia Electronica, Vol 5, Issue 1, 2002.

[3] L. MacDonald, S. Robson: *Polynomial texture mapping and 3D representations*, International Archives of Photogrammetry, Remote Sensing and Spatial Information Sciences, Vol. XXXVIII, Part 5, Commission V Symposium, Newcastle upon Tyne, UK. 2010.

[4] C. Brognara, M. Corsini, M. Dellepiane, A. Giachetti: *Edge Detection on Polynomial Texture Maps*, International Conference on Image Analysis and Processing (ICIAP), 2013.

[5] A. Watt: *3D Computer Graphics*, Addison-Wesley, England, 2000.

[6] O. Demers: *Digital Texturing and Painting*, New Riders, USA, 2002.

[7] G. Earl, K. Martinez, T. Malzbender: *Archaeological applications of polynomial texture mapping: analysis, conservation and representation*, Journal of Archaeological Science 37, pp. 2040-2050, 2010.

[8] J. Meseth, G. Müller, R. Klein: *Reflectance field based real-time, high-quality rendering of bidirectional texture functions*, Computers & Graphics 28, pp. 105–112, 2004.

[9] T. Malzbender, D. Gelb, H. Wolters, B. Zuckerman: *Enhancement of Shape Perception by Surface Reflectance Transformation*, Proceedings of the Vision, Modeling, and Visualisation, Stanford, 2004.

[10] J. Hamiel, J. Yoshida: *Evaluation and Application of Polynomial Texture Mapping (PTM) in the area of Shoe/Tire Impression Evidence*, 2012.

[11] B. Malešević, I. Jovović, M. Makragić, B. Radičić: *A note on solutionsof linear systems*, ISRN Algebra, Vol. 2013, Article ID142124, pp. 1-6, 2013.

[12] I. Jovović, B. Malešević: *A Note on Solutions of the Matrix Equation AXB=C*, Scientific Publications of the State University of Novi Pazar, Ser. A: Appl. Math. Inform. and Mech., Vol 2014 (1), pp. 45-55,2014.

[13] S.L. Campbell, C.D. Meyer: *Generalized inverses of linear transformations*, Society for Industrial and Applied Mathematics, 2009.

[14]  P.S. Stanimirović, G.V. Milovanović: *Simbolička implementacija nelinearne optimizacije*, Elektronski fakultet u Nišu, Edicija monografije, Niš, 2002.

[15] A. Ben-Israel, T.N.E.Greville: *Geralized inverses theory and applications*, Springer, 2003.